\documentclass[prb,twocolumn,aps,superscriptaddress]{revtex4-1}

\usepackage{bm}
\usepackage[colorlinks=true,linkcolor=blue, citecolor=blue]{hyperref}
\usepackage{amsmath}
\usepackage{amssymb}
\usepackage{amsthm}

\usepackage{times}

\usepackage{microtype}
\usepackage{ccaption}
\usepackage{ragged2e}
  \captionnamefont{\bfseries \sffamily}
  \captiontitlefont{\small\sffamily}
  \captiondelim{ \textbar ~ }
  \captionstyle{\justifying}
\usepackage{lineno}

\usepackage{xcolor}
\usepackage{times}

\usepackage{amsfonts}
\usepackage{enumerate}
\usepackage{latexsym}
\usepackage{ifpdf}
\usepackage{graphicx}
\usepackage{color}
\usepackage{makeidx}
\usepackage{ulem}

\expandafter\ifx\csname package@font\endcsname\relax\else
\expandafter\expandafter
\expandafter\usepackage
\expandafter\expandafter
\expandafter{\csname package@font\endcsname}%
\fi
\begin{document}

\title{Anomalous orbital structure in a spinel-perovskite interface $\gamma$-Al$_2$O$_3$/SrTiO$_3$}

\author{Yanwei Cao}
\email{yc003@uark.edu}
\affiliation{Department of Physics, University of Arkansas, Fayetteville, AR, USA}
\author{Xiaoran Liu}
\affiliation{Department of Physics, University of Arkansas, Fayetteville, AR, USA}
\author{P. Shafer}
\affiliation{Advanced Light Source, Lawrence Berkeley National Laboratory, Berkeley, CA, USA}
\author{S. Middey}
\affiliation{Department of Physics, University of Arkansas, Fayetteville, AR, USA}
\author{D. Meyers}
\affiliation{Department of Physics, University of Arkansas, Fayetteville, AR, USA}
\affiliation{Department of Condensed Matter Physics and Materials Science, Brookhaven National Laboratory, Upton, NY, USA}
\author{M. Kareev}
\affiliation{Department of Physics, University of Arkansas, Fayetteville, AR, USA}
\author{Z. Zhong}
\affiliation{Institut f$\ddot{u}$r Theoretische Physik und Astrophysik, Universit$\ddot{a}$t W$\ddot{u}$rzburg, Am Hubland, Germany}
\affiliation{Max-Planck-Institut f$\ddot{u}$r Festk$\ddot{o}$rperforschung, Heisenbergstrasse 1, 70569 Stuttgart, Germany}
\author{J.-W. Kim}
\affiliation{Advanced Photon Source, Argonne National Laboratory, Argonne, IL, USA.}
\author{P. J. Ryan}
\affiliation{Advanced Photon Source, Argonne National Laboratory, Argonne, IL, USA.}
\author{E. Arenholz}
\affiliation{Advanced Light Source, Lawrence Berkeley National Laboratory, Berkeley, CA, USA}
\author{J. Chakhalian}
\affiliation{Department of Physics, University of Arkansas, Fayetteville, AR, USA}
\affiliation{Department of Physics and Astronomy, Rutgers University, Piscataway, NJ, USA}


\begin{abstract} 

\textbf{In all archetypical reported (001)-oriented perovskite heterostructures, it has been deduced that the preferential occupation of two-dimensional electron gases is in-plane $d_\textrm{xy}$ state. In sharp contrast to this, the investigated electronic structure of a spinel-perovskite heterostructure $\gamma$-Al$_2$O$_3$/SrTiO$_3$ by resonant soft X-ray linear dichroism, demonstrates that the preferential occupation is out-of-plane $d_\textrm{xz}$/$d_\textrm{yz}$ states for interfacial electrons. Moreover, the impact of strain further corroborates that this anomalous orbital structure can be linked to the altered crystal  field at the interface  and  symmetry breaking of the interfacial structural units. Our findings provide another interesting route to engineer emergent quantum states with deterministic orbital symmetry.}

\end{abstract}

\maketitle

\begin{figure*}[htp]
\includegraphics[width=0.8\textwidth]{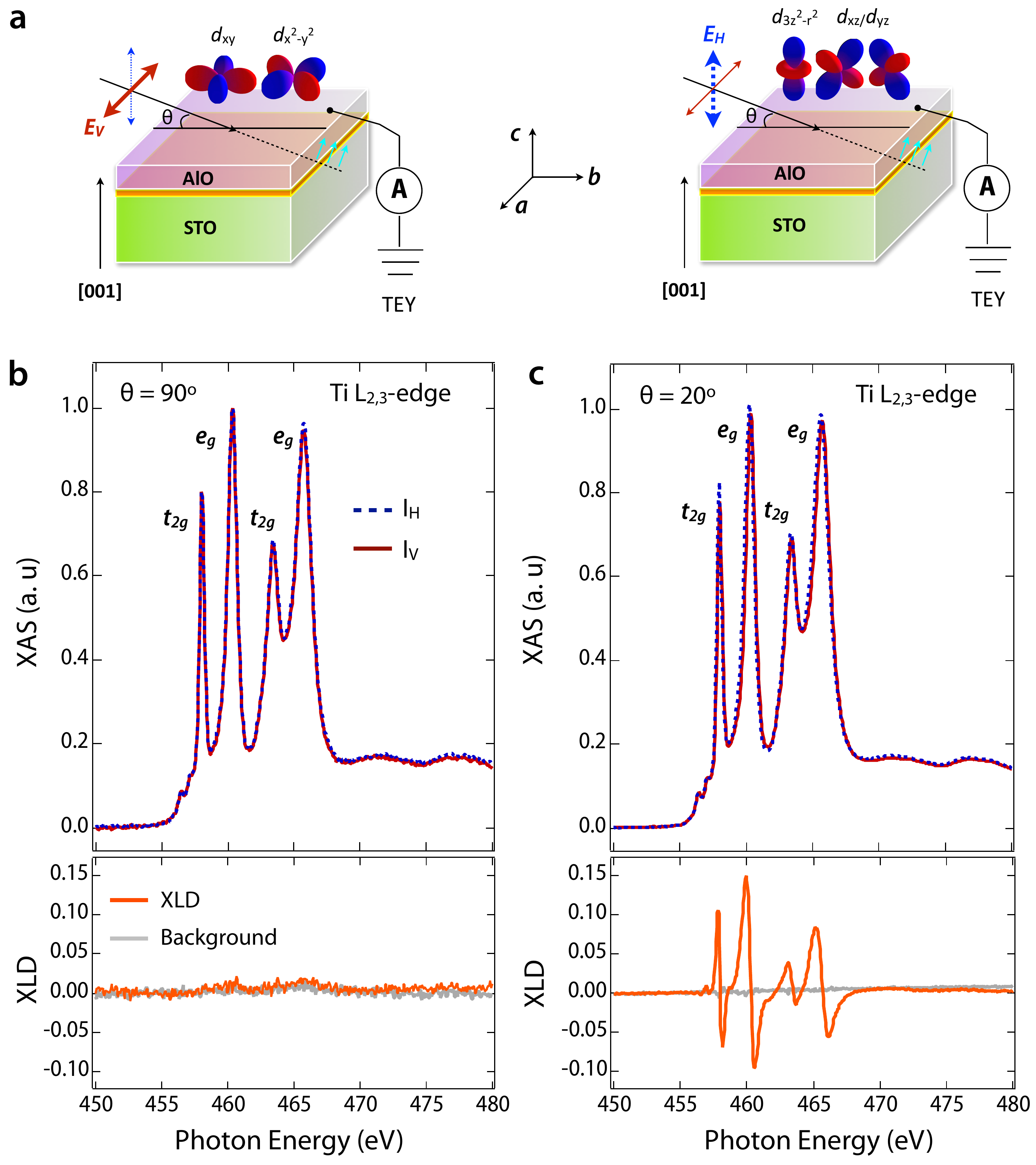}
\caption{\label{} \textbf{Linearly polarized XAS of AlO/STO at Ti L$_{2,3}$-edge.} \textbf{a,} Schematic of experimental setup. In-plane [$I_\textrm{V}$, $E_\textrm{V}$ $\parallel$ $ab$ and $E$ is the linear polarization vector of the photon] and out-of-plane [$I_\textrm{H}$, $\theta$ is the angle between $E_\textrm{H}$ and $c$ ] linearly polarized X-ray were used to measure XAS of AlO/STO (thickness of AlO film is $\sim$ 5.5 unit cells or 4.35 nm) at Ti L$_{2,3}$-edge with total electron yield (TEY, interface sensitive) detection mode at room temperature. The contribution of linearly polarized XAS signal at Ti L$_{2,3}$-edge for t$_\textrm{2g}$ (or e$_\textrm{g}$) band mainly arises from the unoccupied Ti $d_\textrm{xy}$(or $d_{\textrm{x}^2-\textrm{y}^2}$) states by in-plane $I_\textrm{V}$ and $d_\textrm{xz}$/$d_\textrm{yz}$ (or $d_{3\textrm{z}^2-\textrm{r}^2}$) states by out-of-plane $I_\textrm{H}$. Here, the signal of XLD is defined as XLD $=$ [$I_\textrm{H}$ $-$ $I_\textrm{V}$]. \textbf{b,} XAS at Ti L$_{2,3}$-edge with normal incident angle $\theta$ = 90$^{\circ}$. Both $E_\textrm{V}$ $\parallel$ $ab$ and $E_\textrm{H}$ $\parallel$ $ab$. \textbf{c,} XAS at Ti L$_{2,3}$-edge with grazing incident angle $\theta$ = 20$^{\circ}$. As seen in \textbf{a}$, E_\textrm{V}$ $\parallel$ $ab$ whereas $E_\textrm{H}$ $\parallel$ $c$. All collected spectra are repeated more than 6 times.}
\end{figure*}

\begin{figure*}[htp!]
\includegraphics[width=1\textwidth]{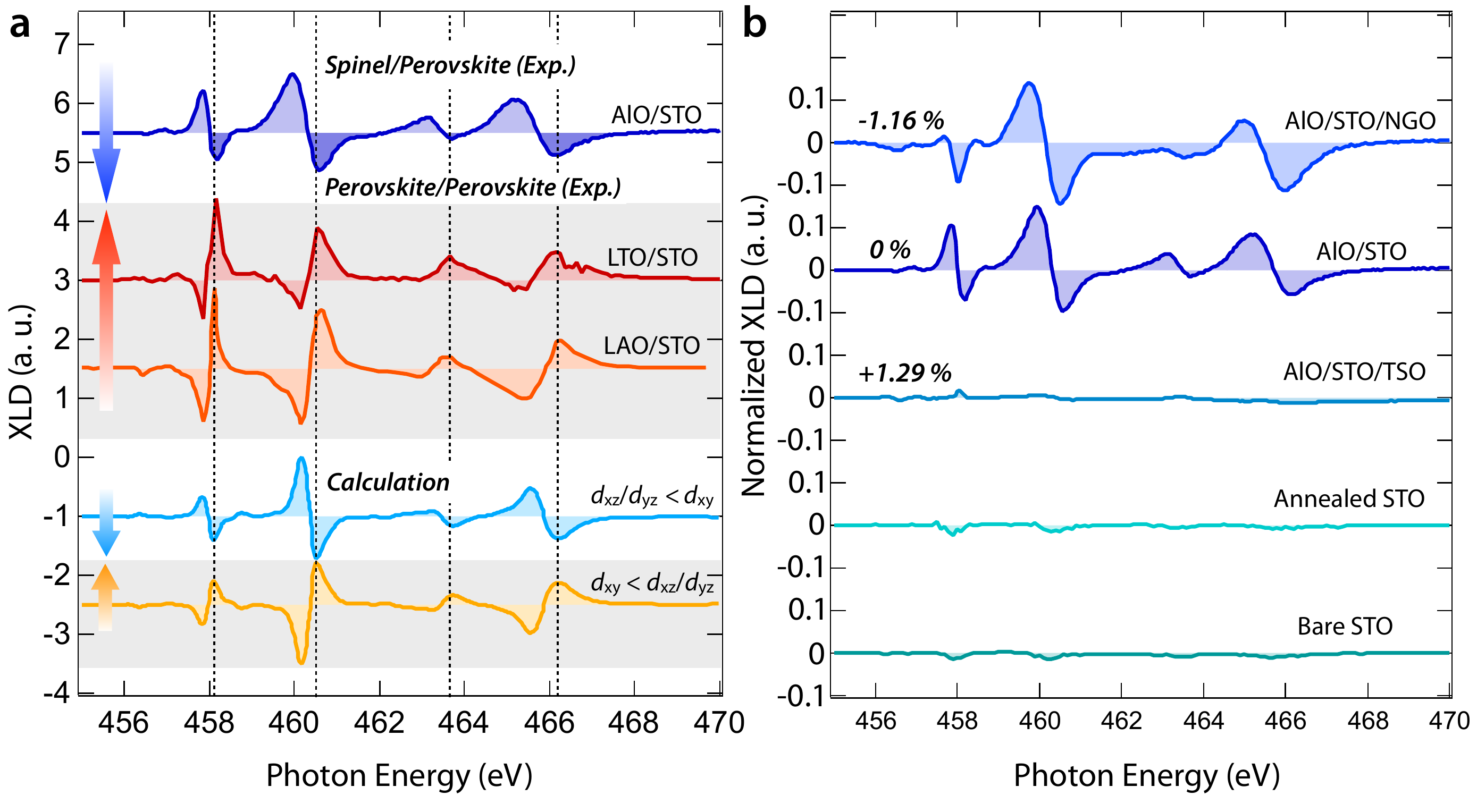}
\caption{\label{} \textbf{Symmetry inversion of XLD spectra in titanate interfaces.} \textbf{a,} The perovskite-perovskite interfaces (i.e. LaAlO$_3$/SrTiO$_3$ and LaTiO$_3$/SrTiO$_3$, red arrow) show negative sign for the first feature (at $\sim$ 457.85 eV), whereas spinel-perovskite heterostructure (AlO/STO, blue arrow) displays positive sign indicating $d_\textrm{xz}$/$d_\textrm{yz}$ is the preferential state of interfacial electrons for the later structure. The spectra of LaAlO$_3$/SrTiO$_3$ was adapted with permission from reference 16. Theoretically, to show the reversed lineshape of XLD for different orbital configurations, the calculation data were adapted with permission from reference 17. \textbf{b,} Strain effects for AlO/STO/NGO and AlO/STO/TSO to XLD signal (compressive strain $\sim$ $-$1.16~$\%$ on NdGaO$_3$ (NGO) and tensile strain $\sim$ $+$ 1.29 $\%$ on TbScO$_3$ (TSO) substrates, respectively; thickness of STO layer is $\sim$ 10 unit cells or 3.9 nm) and effects of oxygen vacancies in annealed STO single crystal to XLD signal. Comparing with the contributions from oxygen vacancies (annealed STO substrate) and bare STO substrate itself, the XLD signal at AlO/STO is robust. (Copyrighted by the American Physical Society.)}
\end{figure*}

\begin{figure*}[htp]
\includegraphics[width=0.85\textwidth]{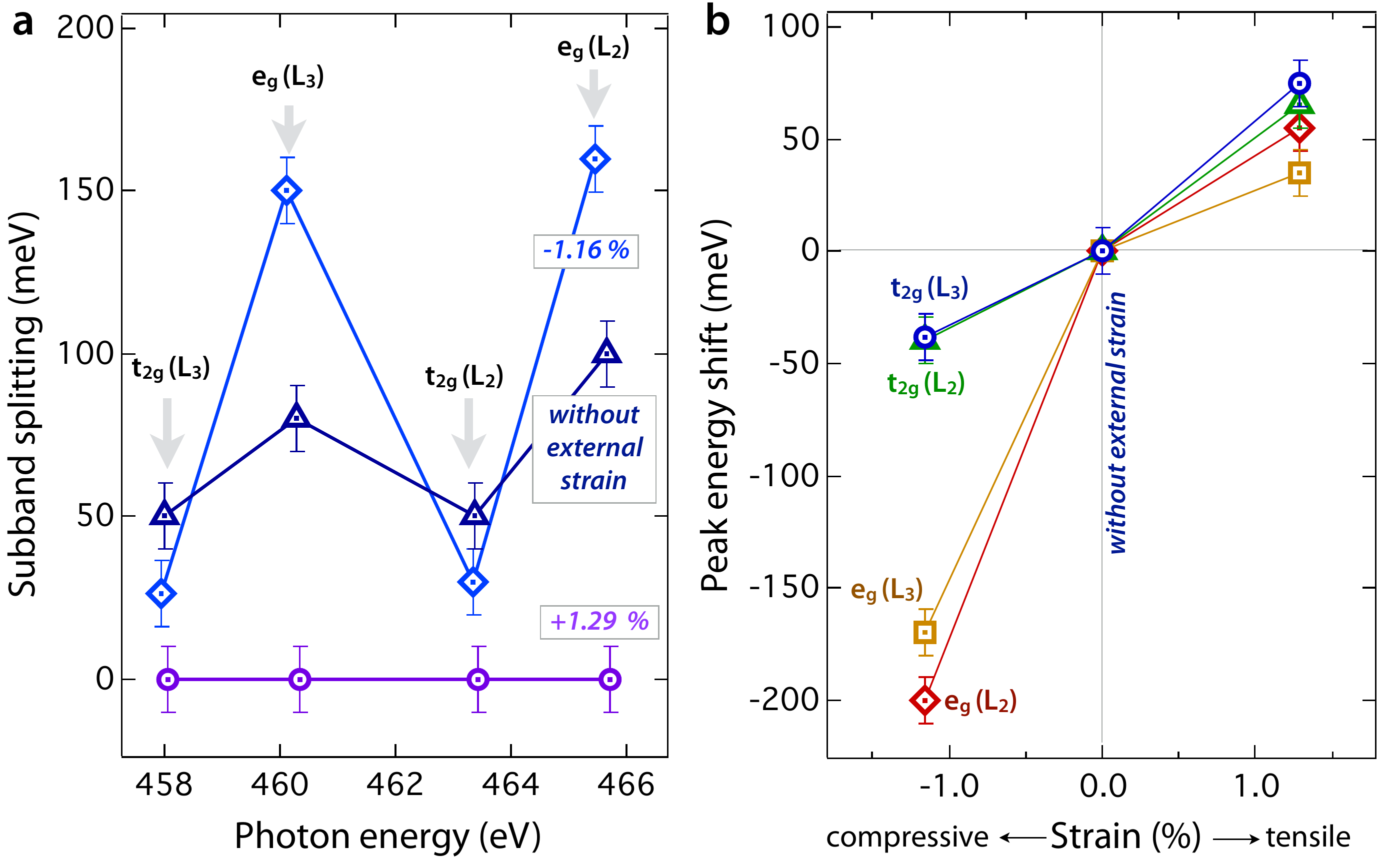}
\caption{\label{} \textbf{Strain effects of Ti 3$d$ subband splitting and peak energy shift.} \textbf{a,} Subband splitting of Ti 3$d$ state at AlO/STO interfaces under compressive strain $\sim$ $-$1.16 $\%$ (by NdGaO$_3$ substrate, NGO), without external strain, and tensile strain $\sim$ $+$1.29 $\%$ (by TbScO$_3$ substrate, TSO). \textbf{b,} Relative peak energy shift (with unpolarized X-rays) of Ti L$_{2,3}$-band in AlO/STO interfaces under compressive strain $\sim$ $-$1.16 $\%$ (negative energy direction) and tensile strain $\sim$ $+$1.29 $\%$ (positive energy direction).}
\end{figure*}

\begin{figure*}[htp!]
\includegraphics[width=0.95\textwidth]{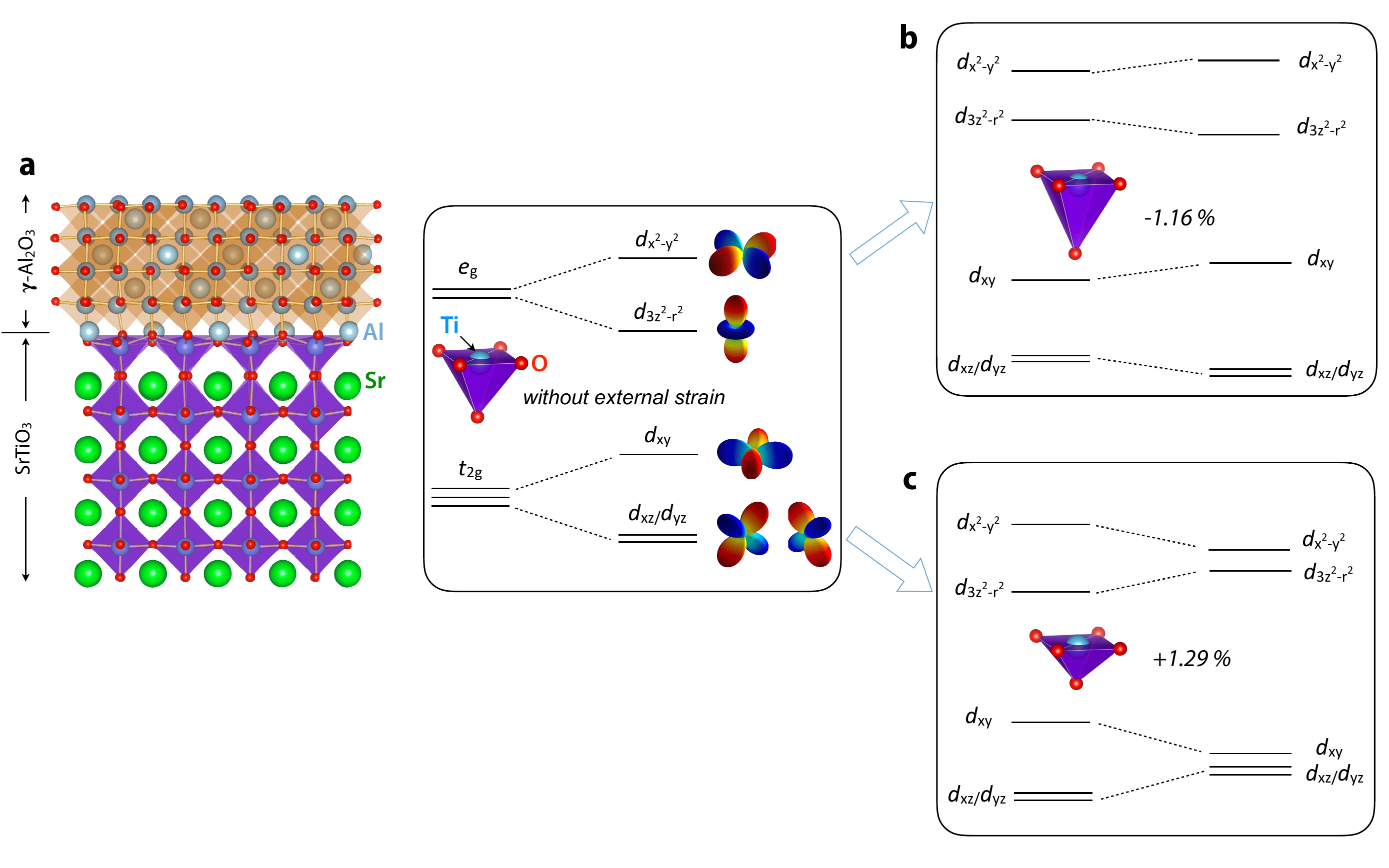}
\caption{\label{} \textbf{Strain effect of crystal field and orbital configuration for AlO/STO interface.} \textbf{a,} Schematic interfacial structure and subband splitting without external strain. \textbf{b,} Compressive strain ($\sim$ $-$1.16 $\%$ by NGO substrate). \textbf{c,} Tensile strain ($\sim$ $+$1.29 $\%$ by TSO substrate).} 
\end{figure*}

\textbf{\large{Introduction}}

Identifying the orbital symmetry of electrons near the Fermi edge is fundamentally important for understanding  phase components of the order parameter \cite{RMP-2000-TS,RMP-2003-Mac,NP-2006-Leg,Nature-2010-Maz,S-2012-Bor}, the coexistence of superconductivity and ferromagnetism \cite{RMP-2014-JC,Nmat-2013-Lee,PRL-2011-Dik,PRL-2012-Mic}, and the high mobility of interfacial conduction electrons \cite{NC-2012-Chen,AM-2014-Chen,NJP-2014-Fete}. For example, at titanates based perovskite-perovskite interfaces, the preferential occupation of an in-plane $d_\textrm{xy}$ state by conduction electrons defines the isotropy of the  Fermi surface \cite{PRL-2009-Sal,AM-2013-Sal,PRL-2013-Sal,PRL-2014-Pes,NC-2015-Her,Nature-2011-SS,NM-2011-Mee,PRL-2014-Wal,PRL-2014-Plu,PRL-2013-Chang,PRB-2013-ZZ,EPL-2012-ZZ,PRL-2016-Cao} of the  two-dimensional electron gas (2DEG). Theoretical calculations have predicted the possibility of out-of-plane $ d_\textrm{xz}$/$d_\textrm{yz}$ orbital symmetry of conduction carriers  in titanates \cite{EPL-2012-ZZ} that can lead to an unusual one-dimensional (1D) electronic structure marked by either two orthogonal 1D bands \cite{Science-1999-Zhou,MRS-2013-Coey,PRB-2016-Yoo} with degenerate $d_\textrm{xz}$/$d_\textrm{yz}$ orbitals or a single 1D band with non-degenerate  $d_\textrm{xz}$ and $d_\textrm{yz}$ orbitals \cite{Nature-2011-SS,PRB-2011-Zhou,NJP-2014-Fete}. Despite extensive experimental search thus far, however,  this anomalous orbital configuration has not been observed  at (001)-oriented interfaces \cite{NC-2015-Her}.

Very recently with the observation of interface enhanced high-temperature superconductivity (60 - 100 K) \cite{CPL-2012-Wang,NM-2015-Ge} and high mobility conduction electrons ($\sim$ 1.4 $\times$ 10$^5$ cm$^2$V$^{-1}$s$^{-1}$) \cite{NC-2012-Chen} in SrTiO$_3$ (STO)-based interfaces, probing the interactions between charge, spin, orbital, and structural degrees of freedom at the interfaces became fundamentally important to understand interface enhanced emergent electronic states. Specifically, the orbital symmetry of conduction carriers is primarily  linked to the symmetry of the superconducting gap \cite{NP-2006-Leg} and the mobility of electrons \cite{NC-2012-Chen,AM-2014-Chen,NJP-2014-Fete}. To this end, the orbital configuration, responsible for the high mobility of electrons in a spinel-perovskite interface (for example $\gamma$-Al$_2$O$_3$/SrTiO$_3$, AlO/STO) \cite{NC-2012-Chen,AM-2014-Chen,JAP-2015-Kor} is still an open question. From the experimental  point of view, surface sensitive angle resolved photoemission spectroscopy (ARPES) with polarized photons is a suitable   probe of  symmetry of the surface electronic structure \cite{Nature-2011-SS}, but it has limited applicability for the electronically active buried interfaces. In contrast  to \ ARPES, interface sensitive linearly polarized x-ray absorption spectroscopy (XAS) has proven to be  a powerful tool to  resolve the orbital  symmetry  \cite{RMP-2014-JC,Science-2007-JC,NC-2016-YC,PRL-2009-Sal,AM-2013-Sal,PRL-2013-Sal,PRL-2014-Pes,Nmat-2013-Lee,PRL-2016-Cao,NC-2015-Her}.

In this work, using the AlO/STO heterostructure as a model system, we report on unique orbital symmetry and orbital occupancy, which is reversed compared to other well-known  2DEGs based on perovskite titanates. Resonant soft X-ray linear dichroism (XLD) studies combined with d.c. transport measurements have confirmed the orbital symmetry inversion driven by the altered crystal  field at
the interface  and  symmetry breaking of the TiO$_6$ octahedral  units. 
\\

\textbf{\large{Results}}

High quality AlO/STO heterostructures  were synthesized by pulsed laser deposition (see Supplementary Fig.~1 and Methods for details). With its spinel structure, bulk $\gamma$-Al$_2$O$_3$ is cubic (space group $Fd\bar{3}m$) with a lattice parameter $a$ = 7.911 \AA \cite{AC-1991-Zhou,JPCM-2010-Jiang}, which is close to twice of the lattice parameter of bulk SrTiO$_3$ $a$ = 3.905 \AA. It is interesting to  note  that  $\gamma$-Al$_2$O$_3$ is generally regarded as a defect spinel Al$_{8/3}$O$_4$ (32 oxygen ions, 64/3 Al cations, and 8/3 vacancies for one unit cell $\gamma$-Al$_2$O$_3$), which has the analogous structure to the conventional spinel MgAl$_2$O$_4$ \cite{JPCM-2010-Jiang}; due to the Al vacancies, polar mismatch may be present at AlO/STO interfaces \cite{PRB-2015-Schutz}. The conductivity of the AlO/STO heterostructures used in this work is in good agreement with the previous reports \cite{NC-2012-Chen,JAP-2015-Kor,JAP-2015-Ngo,PRB-2015-Schutz,APL-2016-Lu}, as shown in Supplementary Fig. 2.

Figure 1 shows the spectra obtained by linearly polarized X-rays at the Ti L$_{2,3}$-edge of AlO/STO heterostructures. 
The features of XAS spectra is the result of  transitions from core levels to unoccupied valence states (e.g. $2p$ $\rightarrow$ $3d$ for Ti L-edge). Due to the crystal field, the Ti 3$d$ state splits into $t_\textrm{2g}$ ($d_\textrm{xy}$, $d_\textrm{xz}$ and $d_\textrm{yz}$) and $e_\textrm{g}$ ($d_{\textrm{x}^2-\textrm{y}^2}$ and $d_{3z^2-r^2}$) subbands with a crystal  field gap as large as $\sim$ 2 eV in the octahedral symmetry \cite{PRL-2009-Sal,AM-2013-Sal,PRL-2013-Sal,PRL-2014-Pes,Nmat-2013-Lee,PRL-2016-Cao,NC-2015-Her}. 
Additionally, the strong spin-orbit interaction induces the splitting of the Ti $2p$ core level into 2$p_{1/2}$ and 2$p_{3/2}$ states. Therefore, four main features are commonly observed in Ti L-edge XAS spectra (see Fig.~1b,c and Supplementary Fig.~3).
With lower crystal symmetry (e. g. tetragonal or orthorhombic symmetry as compared to octahedral symmetry) \cite{Nature-2011-SS}, the degeneracy of $t_\textrm{2g}$ and $e_\textrm{g}$ states can be further lifted, leading to an in-plane $d_\textrm{xy}$ subband with possibly lower energy than the out-of-plane $d_\textrm{xz}$/$d_\textrm{yz}$ subband and available as the lowest energy state at the interface \cite{PRL-2009-Sal,AM-2013-Sal,PRL-2013-Sal,PRL-2014-Pes,Nmat-2013-Lee,PRL-2016-Cao}. To investigate the orbital configuration, XAS with linearly polarized X-rays, used in this work, has been proven to be one of the most powerful available probes applied to various interfaces \cite{PRL-2009-Sal,AM-2013-Sal,PRL-2013-Sal,PRL-2014-Pes,Nmat-2013-Lee,PRL-2016-Cao,NC-2015-Her}. 
The utility of the probe stems from the strong dependence of absorption on the direction of the photon polarization vector ($E$) with respect to  the crystal lattice axis (Fig.~1a);   Thus,  excited by 
linearly polarized X-rays, electronic transitions from Ti core levels to the unoccupied $d$ orbital bands contains important information about the orbital symmetry of those states. In general, when the linear X-ray polarization is oriented along the direction of unoccupied orbital lobes, the contribution of these orbitals to the XAS signal is largest \cite{RMP-2014-JC,Science-2007-JC}. Therefore, the X-ray absorption at the Ti L$_{2,3}$-edge for $E\parallel a$-$b$ and $E\parallel c$ arises mainly from the unoccupied in-plane Ti $d_\textrm{xy}$/$d_{\textrm{x}^2-\textrm{y}^2}$ [$I_\textrm{V}$] and out-of-plane $d_\textrm{xz}$/$d_\textrm{yz}$/$d_{3\textrm{z}^2-\textrm{r}^2}$ [I$_\textrm{H}$] states, respectively. In general, the orbital character of a  subband can be determined from the sign of XLD $\sim$ [$I_\textrm{H}$-$I_\textrm{V}$].

For precise determination of orbital polarization the knowledge of  background can be important. To confirm the low background noise level of XLD and the absence of artifacts,  a normal incidence geometry ($\theta$ = 90$^{\circ}$, Fig.~1b) was utilized. In this geometry, both $E_\textrm{H}$ and $E_\textrm{V}$ are parallel to the interfacial plane and the intensities of linearly polarized XAS should be practically identical for both X-ray polarizations (i.e. [$I_\textrm{H}$-$I_\textrm{V}$]$\sim$
0). As seen in Fig.~1b, no significant XLD signal at Ti L$_{2,3}$-edge is observed in agreement with the expectation \cite{PRL-2014-Pes}. With the sample set at $\theta$ = 20$^{\circ}$, a strong XLD signal appears ($\sim$ 15 $\%$ of XAS, see Fig. 1c) indicating  the splitting of $e_\textrm{g}$ and $t_\textrm{2g}$ subbands with the lineshape that agrees well with the previous measurements and calculations \cite{PRL-2009-Sal,AM-2013-Sal,PRL-2013-Sal,PRL-2014-Pes,Nmat-2013-Lee,PRL-2016-Cao,NC-2015-Her}.

However, as seen in Fig.~2, the XLD spectra for our AlO/STO system is atypical and has the \textit{reverse} XLD lineshape  compared to the results reported for prototypical 2DEGs at titanate interfaces [e.g. LaAlO$_3$/SrTiO$_3$ (LAO/STO) and LaTiO$_3$/SrTiO$_3$ (LTO/STO)] \cite{PRL-2009-Sal,AM-2013-Sal,PRL-2013-Sal,PRL-2014-Pes,Nmat-2013-Lee}. Specifically, for the Ti t$_\textrm{2g}$ state  of the \textit{perovsite-perovskite} interfaces the negative sign of the first  main XLD feature implies that the $d_\textrm{xy}$ subband is the lowest energy state in agreement with reported results \cite{PRL-2009-Sal,AM-2013-Sal,PRL-2013-Sal,PRL-2014-Pes,Nmat-2013-Lee,PRL-2016-Cao}. In sharp contrast to this,  for AlO/STO, the sign of  XLD is \textit{reversed} (see blue and red arrows in Fig.~2a), i.e. the first feature at $\sim$ 457.85 eV has a positive sign whereas the second feature at $\sim$ 458.15 eV is negative, immediately implying that  $d_\textrm{xz}$/$d_\textrm{yz}$ orbitals are the first available states for interfacial electrons. Therefore, the relative energy position of Ti 3$d$ subbands is unusual $d_\textrm{xz}$/$d_\textrm{yz}$ $<$ $d_\textrm{xy}$ $<$ $d_{3\textrm{z}^2-\textrm{r}^2}$ $<$ $d_{\textrm{x}^2-\textrm{y}^2}$. In order to understand 
this anomalous behavior, epitaxial strain was induced by utilizing a large mismatch
between the  substrates and film [i.e. NdGaO$_3$ (NGO)\ substrate for compressive strain $\sim$
$-$1.16~$\%$ and TbScO$_3$ (TSO)\ substrate for tensile strain $\sim$ $+$1.29~$\%$]. As shown in Fig.~2b,
for the AlO/STO heterostructure on NGO  (compressive strain)\  the lineshape
of  XLD is similar to that  observed for AlO/STO except that the very first
 feature (at $\sim$ 457.85 eV) is  suppressed. However, for tensile strain on TSO, surprisingly
almost all the spectral features are killed and no significant XLD signal
is observed.

Next, we quantify the strain effect on the splitting and peak energy shift (see Fig.~3). Generally, the size of the band splitting can be estimated from the peak energy difference of XAS obtained with linear polarized X-rays. First, we analyze the splitting of $e_g$ and $t_\textrm{2g}$ subbands at the AlO/STO interface. As shown in Fig.~3a and Supplementary Fig.~4, a direct comparison of the energy position for XAS with in-plane ($I_\textrm{V}$) and out-of-plane ($I_\textrm{H}$) orientation of the X-ray polarization reveal that the most pronounced XAS feature for $I_\textrm{H}$ is lower in energy than the $I_\textrm{V}$ absorption. For AlO/STO without external strain, it yields  $t_\textrm{2g}$ (L$_{3}$) band splitting $\Delta$ $t_\textrm{2g}$ $\sim$ 50 meV and the $e_\textrm{g} ($L$_{3}$) band splitting $\Delta$ $e_\textrm{g}\sim 80$  meV.  Unexpectedly, as shown in Fig.~3a, under tensile strain ($\sim$ + 1.29~\%,  on TSO substrate) the splitting of both $t_\textrm{2g}$ and $e_\textrm{g}$ bands is suppressed and practically vanished while the splitting is enhanced under compressive strain ($\sim$ $-$1.16~$\%$,  NGO substrate). Compared to the band splitting of AlO/STO without  external strain, the $e_\textrm{g}$ (L$_{3}$) band splitting $\Delta$ $e_\textrm{g}$ under compressive strain ($\sim$ $-$ 1.16~$\%$) increases from $\sim$ 80 meV  to 150 meV, whereas the $t_\textrm{2g}$ (L$_{3}$) band splitting $\Delta$ $t_\textrm{2g}$ $\sim$ 30 meV is only weakly decreased. Besides the subband splitting, strain also alters the peak energy position (see Fig.~3b). Specifically, for tensile strain though the splitting is strongly suppressed (see Fig.~3a and Supplementary Fig.~4) the peak energy moves to the positive direction i.e. higher photon energies by $\sim$ + 75 meV for $t_\textrm{2g}$ (L$_{3}$) and $+$ 35 meV for $e_\textrm{g}$ (L$_{3}$), respectively. On the other hand, as shown in Fig.~3b, under compressive strain with enhanced band splitting, the four main peaks of Ti XAS at L$_{2,3}$-edge shift towards negative direction by about -38 meV for $t_\textrm{2g}$ (L$_{3}$) and $-$ 170 meV for $e_\textrm{g}$ (L$_{3}$), respectively.
\\

\textbf{\large{Discussion}}

Next,
we discuss the atomic structure of AlO/STO interfaces as a key factor to produce the inverse orbital symmetry.   As schematically shown in Supplementary Fig.~5, based on the interfacial atomic structure data \cite{JAP-2015-Kor,APL-2016-Lu},  for the case of $\gamma$-Al$_2$O$_3$ spinel and in contrast to the previously reported \textit{perovskite-perovskite} interfaces, the apical oxygen of Ti-O octahedra is not stable in a \textit{spinel-perovskite} heterostructure. Thus, at the AlO/STO interface a unique Ti-O \textit{pyramid} coordination is formed; in this distorted pyramid-like structure, $d_{\textrm{xz}}$/$d_{\textrm{yz}}$ subband becomes the preferable state for the interfacial electrons (see Supplementary Fig.~5b, c). More importantly, the degenerate $d_{\textrm{xz}}$/$d_{\textrm{yz}}$ subband can be further split due to the cooperative efefct of spin-orbit coupling and crystal field distortion, yielding an energy separation as large as  60-100 meV  \cite{Nature-2011-SS,PRB-2011-Zhou,NJP-2014-Fete}. Therefore, in  contrast with all reported data on the (001)-oriented perovskite interfaces with in-plane $d_{\textrm{xy}}$ subband as the lowest energy state, the $d_{\textrm{xz}}$ or $d_{\textrm{yz}}$ subband becomes the lowest energy state for the case of  \textit{spinel-perovskite} heterojunction. As the consequence of the $d_{\textrm{xz}}$ or $d_{\textrm{yz}}$ orbital character of mobile electrons amplified by the spatial confinement along \textit{z}  \cite{Nat-2002-OA,PRB-2013-You}, and regardless of which $d_{\textrm{xz}}$ or $d_{\textrm{yz}}$ is the preferred state, the forbidden electron hopping along the \textit{y-} (for $d_{\textrm{xz}}$) or \textit{x-} (for $d_{\textrm{yz}}$) direction may result in the emergence  of the extremely anisotropic  \textquotedblleft1D\textquotedblright~electron gas (see Supplementary Fig. 5c).

Furthermore, to understand the impact of epitaxial strain on the XLD signals, we propose a simple model shown  in Fig.~4. As seen, under compressive strain ($\sim$ $-1.16 \%$) the contraction of the in-plane four oxygens together with the elongation of the apical oxygen ion  increases the energy of the in-plane $d_{\textrm{x}^2-\textrm{y}^2}$ and $d_\textrm{xy}$ orbitals, whereas the energy decreases for out-of-plane $d_{3\textrm{z}^2-\textrm{r}^2}$ and $d_\textrm{xz}/d_\textrm{yz}$ orbitals \cite{Kho-2014}. As the result, the energy splitting $\Delta$e$_g$ between $d_{\textrm{x}^2-\textrm{y}^2}$ and $d_{3\textrm{z}^2-\textrm{r}^2}$ orbitals, as well as the splitting $\Delta$t$_{2g}$ between $d_\textrm{xy}$ and $d_\textrm{xz}/d_\textrm{yz}$ orbitals of Ti ions is increased. This model agrees well with the experimental observation that both $\Delta$e$_\textrm{g}$ and $\Delta$t$_\textrm{2g}$  under compressive strain are increased. On the other hand, under tensile strain, the elongation of the in-plane four oxygen ions and the contraction of the apical oxygen ion pulls the Ti ion  inside the pyramid \cite{Kho-2014}, leading to the reversed effect on the Ti 3$d$ orbital sequence. Therefore, the energy splitting within both e$_\textrm{g}$ and t$_\textrm{2g}$ bands is expected to decrease;  the corresponding XLD signal will be  significantly suppressed due to the strain induced degeneracy.     

In conclusion, we have demonstrated that in the \textit{spinel-perovskite} heterostructure - AlO/STO the out-of-plane $d_{\textrm{xz}}$ / $d_{\textrm{yz}}$ states are the lowest lying energy states,  which  is in the sharp contrast to titanate based perovskite-perovskite heterostructures where the in-plane $d_{\textrm{xy}}$ state is always  the ground state of the 2D conduction carriers. Moreover, the impact of strain  corroborates that this unusual orbital configuration  is directly linked with the  altered crystal  field at the interface and  lattice symmetry breaking of the interfacial TiO$_6$ octahedra. Our findings provide another interesting route to engineer  unusual quantum states with deterministic orbital symmetry beyond those attainable in all (001)-oriented perovskite heterojunctions.

\newpage

\textbf{\large{Methods}}

\textbf{Sample synthesis and characterization.} Heterostructures $\gamma$-Al$_2$O$_3$ ($\sim$ 4.35~nm) / SrTiO$_3$ (001), $\gamma$-Al$_2$O$_3$ ($\sim$ 4.35~nm) / SrTiO$_3$ ($\sim$ 3.9~nm) / NdGaO$_3$ (110), and $\gamma$-Al$_2$O$_3$ ($\sim$ 4.35~nm) / SrTiO$_3$ ($\sim$ 3.9~nm) / TbScO$_3$ (110) were layer-by-layer epitaxially grown with pulsed laser deposition (PLD), using a KrF excimer laser operating at $\lambda$~=~248~nm and 2 Hz pulse rate with 2~J/cm$^2$ fluence. The layer-by-layer growth was monitored by \textit{in-situ} reflection-high-energy-electron-diffraction (RHEED). During growth we utilized low oxygen pressure ($\sim$ 7.5$\times$10$^{-5}$ Torr) and the temperature of the substrates was held at $700\,^{\circ}{\rm C}$. After growth, all samples were cooled at about $15\,^{\circ}{\rm C}$/min rate to room temperature keeping oxygen pressure constant. Annealed bulk SrTiO$_3$ was prepared in vacuum ($\sim$ 1$\times$10$^{-6}$ Torr) at $750\,^{\circ}{\rm C}$ for one hour. The lattice parameters of substrates are $a$ = 3.905 \AA~for SrTiO$_3$ (STO); $a$ = 5.43 \AA, $b$ = 5.50 \AA, $c$ = 7.71 \AA~for NdGaO$_3$ (NGO); $a$ = 5.46 \AA, $b$ = 5.72 \AA, $c$ = 7.91 \AA~for TbScO$_3$ (TSO). The sheet-resistances of samples were measured in van-der-Pauw geometry by Physical Properties Measurement System (PPMS, Quantum Design) from 300 to 2~K.  X-ray diffraction was carried out at the 6-ID-B beamline of the Advanced Photon Source at Argonne National Laboratory. 

\textbf{Spectroscopy}. XAS/XLD (at room temperature) at Ti $L_{2,3}$-edge with total electron yield (TEY) detection mode (interface sensitive) were carried out at beamline 4.0.2 of the Advanced Light Source (ALS, Lawrence Berkeley National Laboratory).  In successive scans, spectra were captured with the order of polarization rotation reversed (e. g., horizontal, vertical, vertical, horizontal) so as to eliminate systematic artifacts in the signal that drift with time. The residual artifact intensity is plotted in Fig.~1 and labeled as background.
\\

\textbf {\large{Acknowledgments}}

The authors deeply acknowledge numerous insightful theory discussions with Daniel Khomskii.  J. C. was supported by the Gordon and Betty Moore Foundation EPiQS Initiative through Grant No. GBMF4534. Y. C., S. M., and M. K. were supported by the DOD-ARO under Grant No.0402-17291. X. L. was supported by the Department of Energy Grant No. DE-SC0012375. The Advanced Light Source is supported by the Director, Office of Science, Office of Basic Energy Sciences, of the U.S. Department of Energy under Contract No. DE-AC02-05CH11231. This research used resources of the Advanced Photon Source, a U.S. Department of Energy (DOE) Office of Science User Facility operated for the DOE Office of Science by Argonne National Laboratory under Contract No. DE-AC02-06CH11357.
\\

\textbf {\large{Contributions}}

Y. C. and J. C. conceived and designed the experiments. Y. C., X. L., D. M., P. S. and E. A. acquired the XAS/XLD data. Y. C., X. L., S. M. and D. M. measured the electrical transport. Y. C., X. L., S. M., D. M., J. K. and P. R. measured the X-ray diffraction. M. K. and Y. C. prepared and characterized the samples. Y. C., P. S., E. A. and J. C. analyzed the data. All authors discussed the results. 
\\

\textbf {\large{Competing interests}}

The authors declare no competing financial interests.
\\

\newpage

\end{document}